# Superconducting properties of polycrystalline Nb nanowires templated by carbon nanotubes

A. Rogachev and A. Bezryadin

Department of Physics,

University of Illinois at Urbana-Champaign, Urbana, IL 61801

Abstract

Continuous Nb wires, 7–15 nm in diameter, have been fabricated by sputter-coating single fluorinated carbon nanotubes. Transmission electron microscopy revealed that the wires are polycrystalline, having grain sizes of about 5 nm. The critical current of wires thicker than ~12 nm is very high ($10^7$ A/cm$^2$) and comparable to the expected depairing current. The resistance versus temperature curves measured down to 0.3 K are well described by the Langer-Ambegaokar-McCumber-Halperin (LAMH) theory of thermally activated phase slips. Quantum phase slips are suppressed.

Recently a technique has been developed that employs suspended single-wall carbon nanotubes as templates for material deposition.[1] Because of the small diameter of the nanotubes (1-2 nm) and their chemical stability, it is possible to fabricate ultrathin wires of very different materials. The technique was originally used to obtain sub-10-nm wires of a superconducting amorphous MoGe alloy.[1,2] Later, it was found[3] that continuous nanowires of simple metals – Au, Pd, Fe, Al and Pb – can also be formed if the growth of separate grains is suppressed by prior deposition of a thin (1-2 nm) buffer layer of Ti, a metal with strong chemical bonding to carbon.

In this Letter we show that sub-10 nm thin continuous Nb wires can be fabricated by sputter deposition of Nb over freely suspended carbon nanotubes. Electron microscopy revealed that the structure of the wires is polycrystalline. Nevertheless, the wires of width $w \sim 12$ nm showed the same critical current densities as bulk practical superconductors (Ref. 4, p.372). The resistance versus temperature, $R(T)$, curves measured down to 0.3 K, are well described by the Langer-Ambegaokar-McCumber-Halperin (LAMH) theory of thermally activated phase slips. Quantum phase slips are not observed.

Fluorinated single-wall carbon nanotubes[5,6] employed in the study have approximately $C_2F$ stoichiometry and, unlike ordinary carbon nanotubes, are always insulating. To prepare a sample for transmission electron microscopy (TEM), the nanotubes were first dissolved in isopropanol and then placed on a holey carbon grid. The grid was then dried and placed into a sputtering system equipped with a cryogenic trap. A niobium film of thickness $t \sim 4-10$ nm was then deposited over the samples. Fig.1a shows a high-resolution TEM image of a typical Nb nanowire. The wire is polycrystalline and consists of randomly oriented 3−7-nm grains. Lattice fringes have a spacing of 0.24 nm, which corresponds to (111) lattice planes of Nb. The surface of the wire does not show any crystalline structure due to oxidation. To prevent Nb oxidation, some TEM samples and all measured samples were sputter-coated with a 2-nm Si layer. The image of one of our thinnest wires covered with Si is shown in Fig.1b. The observed width variation in Nb wires ($\pm 2$ nm) is larger and has a longer characteristic wavelength along the wire than that found in amorphous MoGe wires.[1] None of about 20 Nb nanowires studied with TEM showed any interruptions in the Nb core, such as those found in Al or Au wires in Ref. 3. All Nb wires appeared continuous.

Samples for transport measurements were fabricated following the protocols of Ref.1. Measurements of resistance (Fig.2a) were made in vacuum with small bias current (0.5-10 nA). The parameters of Nb nanowires are given in a caption to the Fig.2. Some of our samples have been remeasured after a storage over six months. The samples Nb5 and Nb6 did not changed. The samples which changed notably are labeled with a letter "a" added to a original name. One sample of these samples, Nb8a, showed an insulating behavior. For each curve in Fig.2a, a resistance drop at higher temperature corresponds to a superconducting transition in Nb film electrodes connected in series with the nanowire. The resistance measured immediately below the film transition is taken to be the normal-state resistance of the wire, $R_N$. To estimate bulk resistivity, *r,* of Nb in the wires we plot the data for as prepared samples in the form $L/(tR_N)$ versus $w$ (Fig.2b), where $L$ is the length of a wire, $t$ is the thickness of the deposited Nb film and $w$ is the width seen in the scanning electron microscope (SEM) image. Similar to the case of MoGe wires,[1] Nb data points follow a straight line with the slope of 1/*r* in such a representation. The intersection with the horizontal axis accounts for the errors in the width determination that might be caused by (i) SEM limited resolution; (ii) conductively dead layers at the interfaces of the Nb core with the nanotube and Si; (iii) non-uniformity of Nb coverage along the wire; (iv) partial oxidation of Nb core. From the slope of the linear fit we find $r \approx 30$ µΩ cm. Then from the value of $r\ell = 3.72 \times 10^{-6}$ µΩ cm known for Nb,[7] we determine a mean free path $\ell \approx 1.2$ nm. These values are in an agreement with those reported for thin polycrystalline Nb films.[7,8]

The second transition seen for the Nb1-Nb6 samples (Fig.2a) corresponds to superconducting transition in the wire itself. In a one-dimensional superconductor, finite resistance below $T_c$ arises because of thermally activated phase slips. The theory of this process was developed by Langer, Ambegaokar, McCamber and Halperin (LAMH).[9] The resistance is given by the equation

$$R_{LMAH}(T) = \frac{\pi\hbar^2}{2e^2 kT t_{GL}} \frac{L}{x(T)} \left(\frac{\Delta F}{kT}\right)^{-1/2} \exp\left(-\frac{\Delta F}{kT}\right), \quad (1)$$

where we use $\Delta F(T) \approx 0.83(L/x(0))(R_q/R_N)kT_C(1-T/T_C)^{3/2}$ as the energy barrier (see Eqs.4 and 4a in Ref. 2), $t_{GL} = \pi\hbar/8k(T_C - T)$ is the Ginzburg-Landau relaxation time, $R_Q = h/4e^2 = 6.45$ kΩ is quantum resistance for Cooper pairs and $x(T) = x(0)(1-T/T_C)^{-1/2}$ is the coherence length. The expression $R(T) = (R_{LAMH}^{-1} + R_N^{-1})^{-1}$ was used for fitting experimental data, where $R_N$ is the normal state resistance of the wire. We use two fitting parameters: the coherence length $x(0)$ and $T_C$. The best fits (solid lines) are shown in Fig.3. The experimental data follows very well the LAMH dependence in samples Nb1-Nb3, Nb5 and Nb4a (not shown). Some non-negligible deviation from LAMH is observed in the sample Nb6. The sample Nb8a is insulating and can not be described by LAMH. The extracted coherence length for samples Nb1 and Nb2 ($x(0) \approx 8$ nm) is comparable to $x(0) \approx 7$ nm found in polycrystalline Nb films.[10] It also agrees well with an estimate $x(0) \approx (x_0 \ell)^{1/2}$=6.9 nm, where $x_0$=40 nm is the coherence length for clean Nb, and $\ell \approx 1.2$ nm. For the samples Nb3, Nb5 and Nb6, the best fits corresponds to larger values of $x(0)$ (16-18 nm). Partially, the increase of $x(0)$ is due to expected[11] suppression of $T_C$, since $x(0) \sim (1/T_c)^{1/2}$ in the dirty limit. The deviation from the LAMH dependence found in Nb6 most probably is due to fluctuations in the width of the wire and corresponding fluctuations of the local $T_C$ values[11]. Note that the samples Nb6 and Nb8a have $R_N > R_Q$ so localization effects could be significant in these samples.

Deviations from LAMH have been found in many previous experiments[12,13,14,2]. A contribution due to macroscopic quantum tunneling (MQT) [12,2] of phase slips was then introduced to explain characteristic low-temperature resistance tails observed in some cases. The $R_{MQT}$ was suggested to follow the from of Eq.1 with $kT$ being replaced with $\hbar/t_{GL}$ (Ref.12, 2). It is also expected to dominate over the thermal resistance below $\sim T_c/2$.[15] To see whether the MQT is present in Nb nanowires we calculate the $R_{MQT}$, using Eq.2 in Ref.2, with generic numerical factors $a$=1 and $B$=1 and $a$=1.3 and $B$=7.3 found for MoGe wires in Ref.2. The QPS contribution is computed using the values of $x(0)$ and $T_C$ extracted from the LAMH fits. We show the total expected resistance, $R(T)=[R_N^{-1}+(R_{LAMH}+R_{MQT})^{-1}]^{-1}$, as dashed lines (for $a$=1 and $B$=1) and dotted lines (for $a$=1.3 and $B$=7.3) in Fig.3. It is clear that for both cases the MQT model strongly deviates from the experimental curves for samples Nb3, Nb5 and Nb6, so quantum phase slips appear to be suppressed in Nb nanowires. On the other hand, it is of course possible to increase the $a$ value and shift the MQT effect to lower temperatures. In our experiments, if $a$>4, the MQT contribution could not be resolved.

We have performed voltage versus current, $V(I)$, measurements for some Nb wires. Below we present data for a representative sample Nb2. With decreasing temperature, $V(I)$ undergoes the following transformations. Slightly below $T_c$ and at high bias currents (Fig. 4a), $V(I)$ follows the exponential dependence $V(I) \sim \exp(I/I_0)$. The experimental coefficient $I_0 \approx 0.09$ mA is close to the LAMH prediction $I_0 = 4ekT/h \approx 0.06$ mA (Ref. 16, p.291). At lower $T$, the resistance becomes immeasurably low until the critical current is reached. At the critical current, the sample shows a few voltage steps (Fig. 4b) and then, at higher bias currents, enters a dissipative regime with a linear $V(I)$ dependence. Each new step in the $V(I)$ curve is due to appearance of a new phase slip center (PSC) in the wire.[16] Such multi-step $V(I)$ curves indicate that the dissipative size of a single PCS is less than the length of the wire. When the temperature decreases further, the steps merge, probably due to a synchronization of PSCs.[17] At lowest temperatures we always see only one critical current, i.e. one large step (Fig. 4c). The hysteresis (Fig. 4c) can be associated either with heating or with the dynamical effects described in Ref.18. The heating seems less probable since $V(I)$ curves above the critical current are perfectly linear, exhibit a non-zero offset current (i.e. do not extrapolate to the origin), and are parallel to the normal-state dependence ($V=IR_N$). This suggests that a nonzero average supercurrent does flow through the wire, even in the dissipative regime.

We now compare the critical current $I_c(0)$ of the sample Nb2 extrapolated to $T=0$ K and the depairing critical current, $I_{dp}$, that can be calculated as $I_{dp} =(92\textbf{\textit{m}}A)LT_c/R_N\textbf{\textit{x}}(0)$ (Eq.4 in Ref.15), using parameters obtained from the LAMH fit. This expression gives $I_{dp}\approx 12$ $\textbf{\textit{m}}A$, reasonably close to experimental value of $I_c(0)\approx 8$ $\textbf{\textit{m}}A$, thus confirming the consistency of our $I_c$ and $R(T)$ measurements. Another approach is to use parameters for bulk clean Nb and the expression for the Ginzburg-Landau depairing critical current density, $J_{dp}= (2/3)^{3/2}(H_C(0)/\textbf{\textit{l}})$. Since $H_C(0) \sim T_C$ and $T_C=9.2$ K for bulk Nb, $H_C(0)$ in the Nb2 wire, which has $T_C \approx 5.6$ K, should be reduced by a factor of 0.6, compared with the bulk value 0.2 T. The penetration depth is given by the expression $\textbf{\textit{l}} = \textbf{\textit{l}}_L\left(1+\textbf{\textit{x}}_0/\ell\right)^{1/2}$. Then, with $\textbf{\textit{l}}_L$=40 nm and $\textbf{\textit{x}}_0$=40 nm for clean Nb and $\ell$=1.2 nm, we have $\textbf{\textit{l}} \approx 235$ nm and $J_{dp} \approx 2\times 10^7$ A/cm$^2$. We can estimate the experimental critical current density of the Nb2 wire as $J_c(0)=I_c(0)R_N/\textbf{\textit{r}}L$. With $\textbf{\textit{r}}$=30 μΩ cm this gives $J_c(0) \approx 10^7$ A/cm$^2$. The values of $J_c(0)$ and $J_{dp}$ are quite close to each other. This demonstrates that thicker polycrystalline Nb wires deposited on carbon nanotubes do not have weak links, and the grain boundaries seen in the TEM image (Fig. 1a) do not produce tunneling barriers for supercurrent.

This work was supported by NSF CAREER grant DMR-01-34770 and by the Alfred P. Sloan foundation fellowship. Part of the work was carried out in the Center for Microanalysis of Materials, University of Illinois, which is partially supported by the U.S. Department of Energy under grant DEFG02-91-ER45439. The authors thank J. Margrave for providing them with fluorinated carbon nanotubes and V. Ryazanov and A. Bollinger for useful suggestions.

# Figure captions

*Figure 1*

(a) A high-resolution TEM image of a Nb nanowire fabricated by deposition of a 6-nm Nb film over a single-wall carbon nanotube. (b) Image of one of the thinnest Nb nanowire (the thickness is 4 nm) covered with a protective layer of Si (2 nm) visible as light layer at the surface.

*Figure 2*

(a) Temperature dependence of the resistance of Nb nanowires. The samples Nb1, Nb2, Nb3, Nb3a, Nb4, Nb4a, Nb5, Nb6, Nb7, Nb8 have the following parameters. Normal state resistances (k$\Omega$) are $R_N$=0.47, 0.65, 1.61, 1.8, 2.35, 2.73, 4.25, 9.5, 15.7 and 47.5 respectively. Lengths (nm) are $L$=137, 120, 172, 172, 177, 177, 110, 113, 196 and 235 respectively. (b) A value of $L/(R_N t)$ plotted versus wire width that was determined from SEM images. The solid line shows the best linear fit.

*Figure 3*

Temperature dependence of the resistance of superconducting Nb nanowires. Solid lines show the fits to the LAMH theory. The samples Nb1, Nb2, Nb3, Nb5, Nb6 have the following fitting parameters. Transition temperatures (K) are $T_C$=5.8, 5.6, 2.7, 2.5 and 1.9 K respectively. Coherence lengths (nm) are $x(0)$ =8.5, 8.1, 18, 16 and 16.5 respectively. The dashed lines are theoretical curves that include the contribution of quantum phase slips into the wire resistance (Ref.2), with generic factors $a$=1 and $B$=1. The dotted lines are computed with $a$=1.3 and $B$=7.2.

*Figure 4*

Voltage versus current dependence for sample Nb2. (a) A $V(I)$ dependence measured at $T$=4.8 K (open circles) in a log-linear representation. The straight solid line is a guide to the eye; (b) Voltage-current curves for different temperatures close to $T_c$. The corresponding temperatures from left to right are 4.76, 4.68, 4.58, 4.47, and 4.53 K. The stepwise behavior corresponds to phase slip centers. (c) Hysteretic $V(I)$ variation at the lowest temperature, $T$=1.58 K. The critical current, $I_c$, the retrapping current, $I_r$, and the offset current, $I_s$, are indicated by arrows.

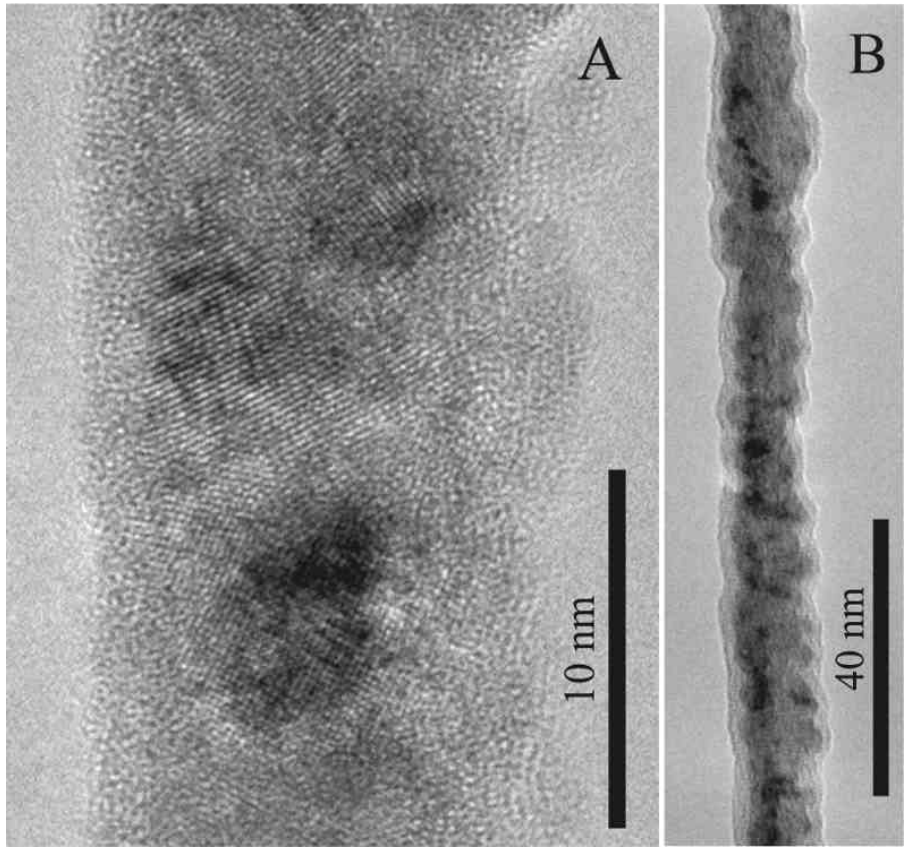

**Figure 1**

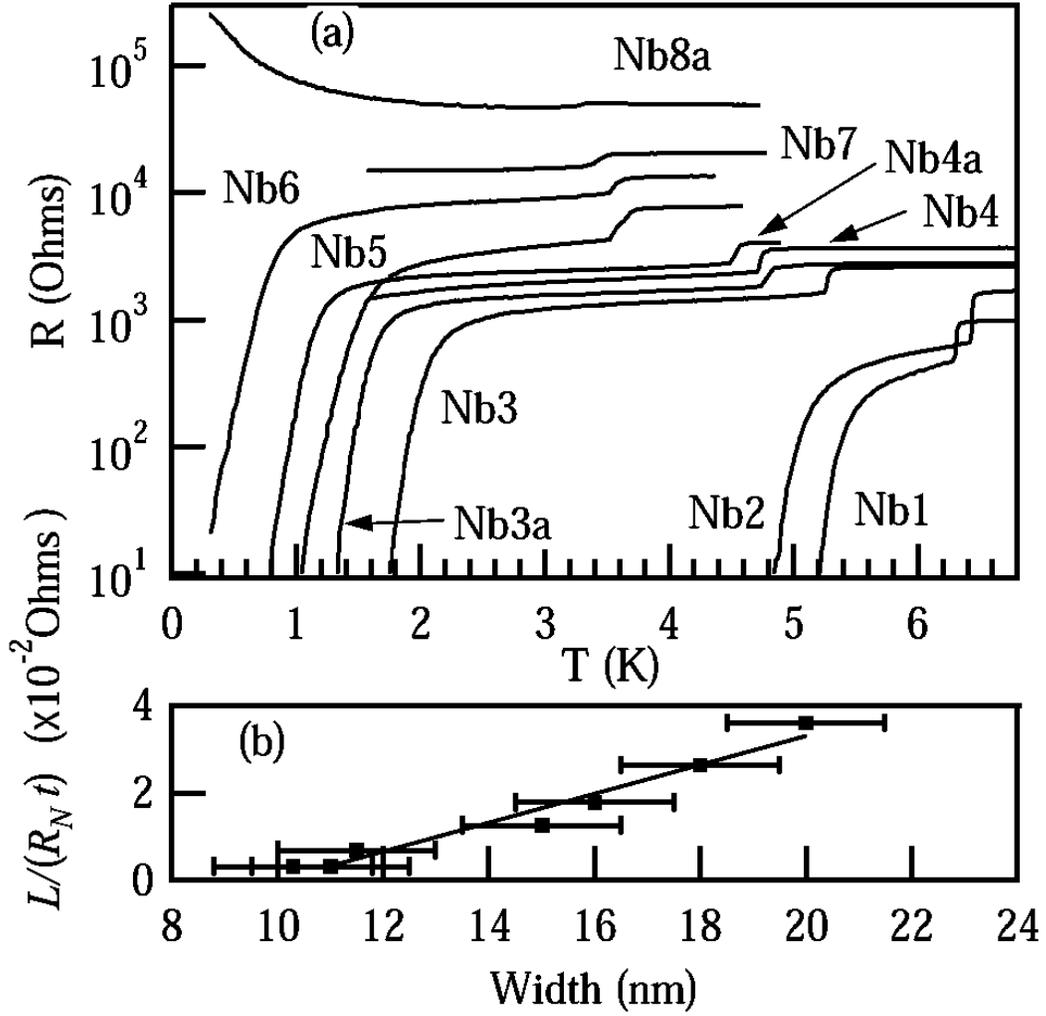

**Figure 2**

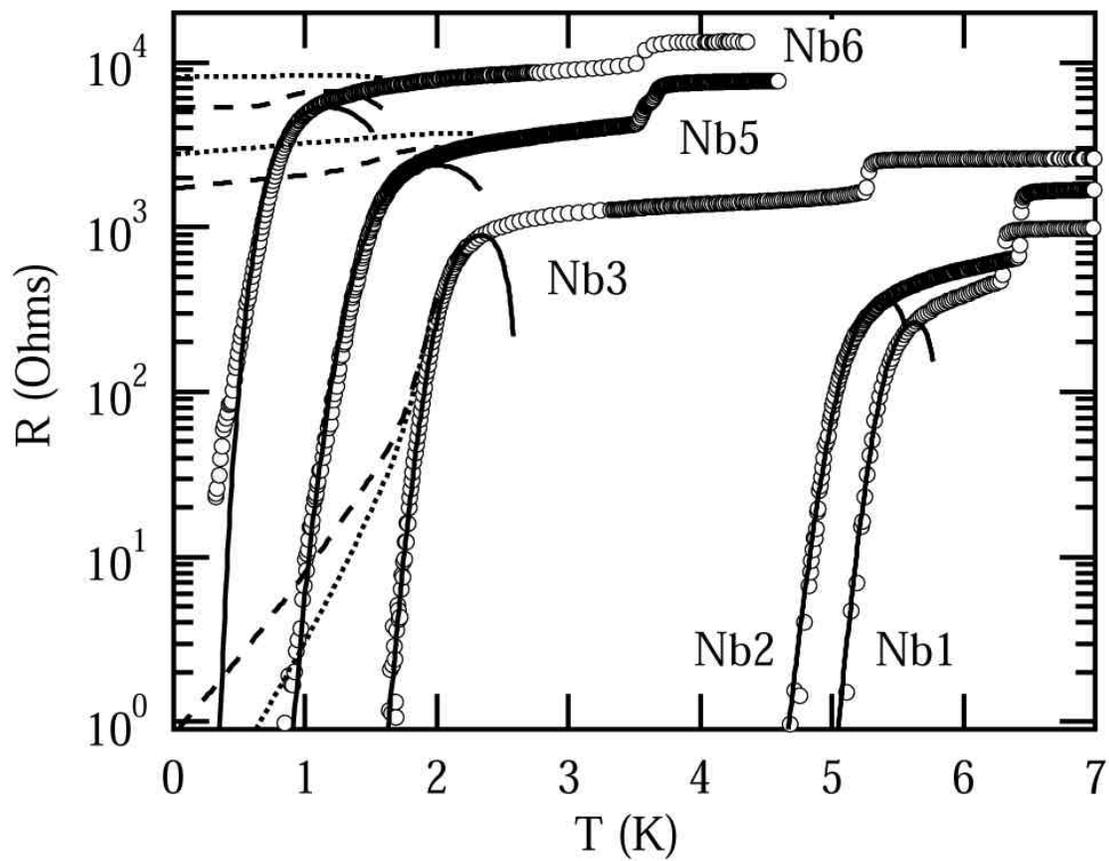

**Figure 3**

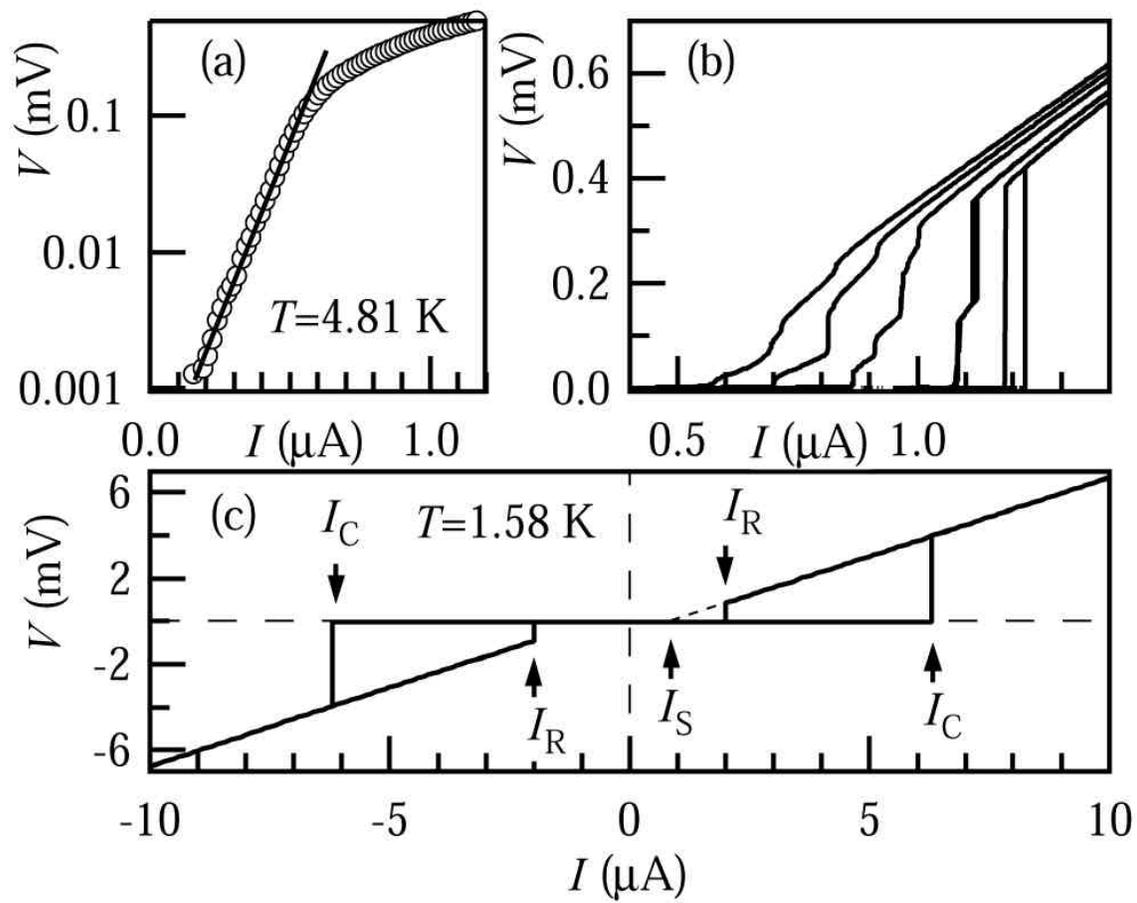

**Figure 4**